\documentclass[reprint,twocolumn,showpacs,amsmath,amssymb,superscriptaddress,aps,prl]{revtex4-2}
\usepackage{amsmath}
\usepackage{amsfonts}
\usepackage{amssymb}
\usepackage{graphicx}
\usepackage{epsfig}
\usepackage{color}
\usepackage{empheq}
\usepackage{lineno}
\usepackage{hyperref}

\newcommand{\RR}{\right}
\newcommand{\LL}{\left}
\newcommand{\m}{\textrm}
\newcommand{\dg}{\dagger}
\newcommand{\eref}[1]{Eq.~(\ref{#1})}
\newcommand{\fref}[1]{Fig.~\ref{#1}}
\newcommand{\puoli}{\frac{1}{2}}
\newcommand{\rtHz}{\sqrt{\mathrm{Hz}}}

\begin{document}

\title{Quantum-enabled measurement of gravitational forces at the milligram scale}

\author{Matthew Herbst}
\affiliation{Department of Applied Physics, Aalto University, P.O. Box 15100, FI-00076 AALTO, Finland}

\author{Ewa Rej}
\affiliation{Department of Applied Physics, Aalto University, P.O. Box 15100, FI-00076 AALTO, Finland}

\author{Joe Depellette}
\affiliation{Department of Applied Physics, Aalto University, P.O. Box 15100, FI-00076 AALTO, Finland}

\author{Richa Cutting}
\affiliation{Department of Applied Physics, Aalto University, P.O. Box 15100, FI-00076 AALTO, Finland}

\author{Jinhoon Jeong}
\affiliation{Department of Applied Physics, Aalto University, P.O. Box 15100, FI-00076 AALTO, Finland}

\author{Mika A. Sillanpää}
\affiliation{Department of Applied Physics, Aalto University, P.O. Box 15100, FI-00076 AALTO, Finland}


\begin{abstract}
\textbf{Gravity is by far the weakest among all known interactions. While it dominates the large-scale structure of the universe, gravitational interactions between small masses in the microscopic range remain underexplored because they are too weak to be experimentally accessible. The interface between gravity, described by general relativity, and quantum mechanics, which describes the microscopic world, is of particular interest. From a theory standpoint, these two major descriptions of nature have not been combined in a satisfactory manner \cite{Penrose2014}. Technology is gradually becoming mature enough such that probing gravity at the microscopic scale, where quantum mechanics and gravity potentially meet, is becoming accessible \cite{pikovski_probing_2012,Bose2017qgrav,Vedral2017qgrav}. Explorations towards genuine quantum gravity have to be carried out under conditions where the gravitating masses possess non-classical properties, which necessitates ultimate protection against disturbances, such as ultralow temperatures and a small range of motion. Here, we present an experiment at millikelvin temperatures, where the gravity due to a micron-range coherent displacement of an $\mathbf{87}\,\m{mg}$ source mass imprints on the motion of an oscillating membrane loaded with a $\mathbf{1.3}\,\m{mg}$ test mass. Vibrations of the membrane are measured in the microwave optomechanical scheme \cite{teufel_sideband_2011}, which is a key setup for investigating non-classical motional states of mechanical oscillators \cite{aspelmeyer_cavity_2014,mercier_de_lepinay_quantum_2021,Teufel2021entangle}. When considering the combination of total source and test masses, our setup presents two orders of magnitude progress in sensing gravity at the small scale \cite{mitrofanov_experimental_1988,Tan2016torsion,lee_new_2020,westphal_measurement_2021}, as well as a $\mathbf{35}\,\m{dB}$ improvement of gravitational sensitivity at kHz frequencies or above. Our setup offers a plausible approach \cite{liu_gravitational_2021} for studying gravitational interactions between masses prepared in non-classical motional states.}
\end{abstract}

\maketitle

It is still unclear whether gravity is a quantum-mechanical entity, or whether it is somehow disconnected from the rest of the universe that obeys quantum mechanics \cite{Penrose2014,Oppenheim2023PRX}. Finding a unified theory of quantum gravity is hindered by a lack of data, with the energy, length, and time scales at which quantum mechanics and gravity may plausibly meet being experimentally inaccessible in the foreseeable future. However, it has been recently realized that the gravity-quantum interface may be probed indirectly, by devising an experiment in which two small masses in non-classical states interact gravitationally \cite{Bose2017qgrav,Paternostro2020Gentangle,liu_gravitational_2021}. 
To reach this goal, we must push gravity experiments to smaller masses, while enabling quantum control for ever larger objects until gravity and quantum scales meet. 

The first part of this challenge can trace its roots back to Cavendish's torsion pendulum experiment in the 1700's
\cite{cavendish_xxi_1798}. Since then, these devices have been scaled down well into the gram-scale \cite{gillies_attracting_2014,mitrofanov_experimental_1988, lee_new_2020,ritter_experimental_1990} and many other experimental platforms have been developed. Prominent examples are laboratory-sized source masses influencing levitating test masses \cite{yin_experiments_2022, fuchs_measuring_2024} or trapped atoms \cite{rosi_precision_2014, panda_measuring_2024}. 
Torsion pendulums have remained at the forefront of small-mass experiments, with a recent experiment detecting gravity from a set of sixteen $7.6\,\m{mg}$ source masses \cite{Yang2026smallmass}. The record for the two lightest gravitationally coupled objects, however, remains a 2021 experiment using $91\,\m{mg}$ and $92\,\m{mg}$ gold spheres \cite{westphal_measurement_2021}.

The second aspect is to prepare objects in non-classical center-of-mass (COM) states. The largest pair of masses with entangled COM states are micromechanical oscillators approaching $0.1\,\m{ng}$ \cite{ockeloen-korppi_stabilized_2018, riedinger_remote_2018}. In these cavity optomechanical or electromechanical systems, mechanical oscillators are parametrically coupled to electromagnetic fields inside a cavity \cite{aspelmeyer_cavity_2014}. By cooling a mechanical oscillator with resonance frequency $f_\m m$ down to below the temperature $h f_\m m / k_\m B$ (with Planck constant $h$ and Boltzmann constant $k_\m B$), the motional ground state is reached.
Encouragingly, systems with much larger masses, which are the most promising for gravity experiments, have been prepared close to the ground state \cite{LIGO2021_10kg,Chu2023_500ug} or in genuinely non-classical states \cite{Chu2023cat}.

\begin{figure*}[t]
    \centering
    \includegraphics[width=0.9\textwidth]{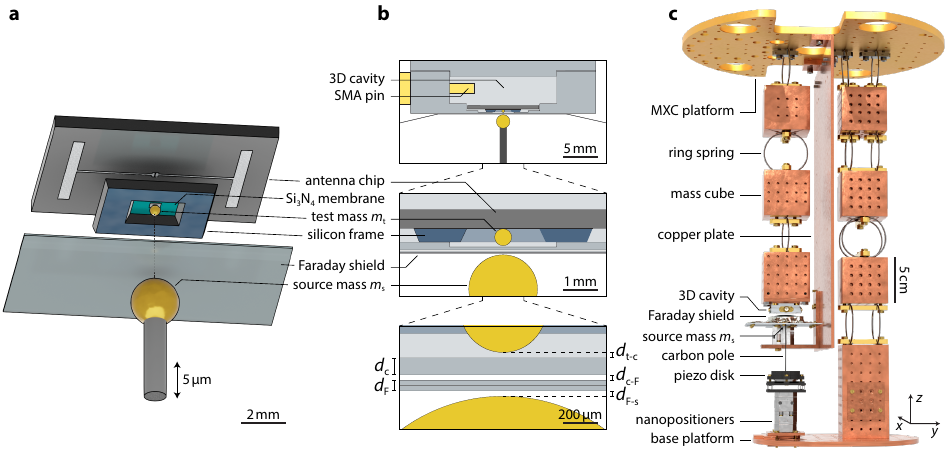}
    \caption{\textbf{Experimental setup.} 
    \textbf a: Exploded view  of the main components of the assembly (from top to bottom): antenna chip for microwave optomechanical measurement, the mechanical oscillator formed by silicon nitride membrane (cyan) embedded with the test mass, the Faraday shield, and the source mass. The 3D cavity with its wall between the Faraday shield and the test mass is not shown for clarity.
    \textbf b: To-scale cross-sectional view of the setup at three levels of zoom. Top: The flip-chip is glued in the bottom of the superconducting 3D cavity made of aluminum. The microwave measurements are carried out via the SMA coupler. The source mass is mounted to a carbon fiber pole. Center and bottom: The test mass is spaced by $d_\m{t-c} \simeq 30\,\m{\textmu m}$ surface distance from the inner cavity wall that is $d_\m{c} \simeq 100\,\m{\textmu m}$ thick. The movable Faraday shield that consists of two separate aluminum layers is $d_\m{F} \simeq 60\,\m{\textmu m}$ thick. The shield-cavity surface distance  $d_\m{c-F}$ and the source mass-shield distance $d_\m{F-s}$ are independently variable.
    \textbf c: Large-scale cryogenic set-up. For clarity, we have omitted all electrical cabling and thermal links. The mass-spring assembly on the left suspends the 3D cavity from the mixing chamber (MXC) platform of a dilution refrigerator. A separate suspension chain on the right supports the $xyz$-nanopositioners and the driving piezo, to which the carbon pole with source mass is glued. The Faraday shield is attached to another nanopositioner which is mounted on a large copper plate running through the center of the set-up.
    }
    \label{fig:experiment}
\end{figure*}


Here, we address the challenges of measuring gravity between two milligram masses in a new experiment that is enabled by the cavity-optomechanical techniques at the forefront of quantum COM control. A vibrating gold sphere acts as a gravity source mass  $m_\m s = 86.6\,\m{mg}$. Its motion $\delta z_\m s(t) = z_\m s \cos \omega_\m d t$ modulates its COM distance $d_\m{com}(t) = d_\m{com,0} + \delta z_\m s(t)$ to a second gold sphere acting as a test mass $m_\m t = 1.26\,\m{mg}$. 
The test mass experiences a dynamic Newtonian force 
$\delta F_\m G(t) = - F_\m G \cos \omega_\m d t$ 
with an amplitude 
$F_\m G = \frac{2 G m_\m t m_\m s}{d_\m{com,0}^3} z_\m s$ and gravitational constant $G = 6.67\cdot10^{-11}\,\m m^3\m{kg}^{-1}\m s^{-2}$. 
With the tiny masses and only micron-sized displacements involved here, this force lies in the low attonewton range. 

\textbf{Cryogenic apparatus.} 
The key component to the experiment
is a superconducting microwave opto/electromechanical device housed in a cryogen-free dilution refrigerator. It consists of a freestanding mm-size silicon nitride membrane, which is loaded with the $r_\m t = 250\,\m{\textmu m}$ radius gold sphere acting as the test mass $m_\m t$ as illustrated in \fref{fig:experiment}\textbf{a}.
Owing to the high pre-stress of the membrane, the frequency of the lowest drum-like mode, where the mass moves up and down at the antinode of the mode, is above $2\,\m{kHz}$, significantly higher than the millihertz frequencies used by torsion-pendulum based experiments \cite{westphal_measurement_2021, Yang2026smallmass}. 
The oscillator state is accessed via microwave antennas inside a superconducting 3D cavity \cite{yuan_large_2015,noguchi_ground_2016,rej_near-ground_2024}, which allow for both detection and manipulation of the vibrational state.


\begin{figure*}[t]
    \centering
    \includegraphics[width=0.95\textwidth]{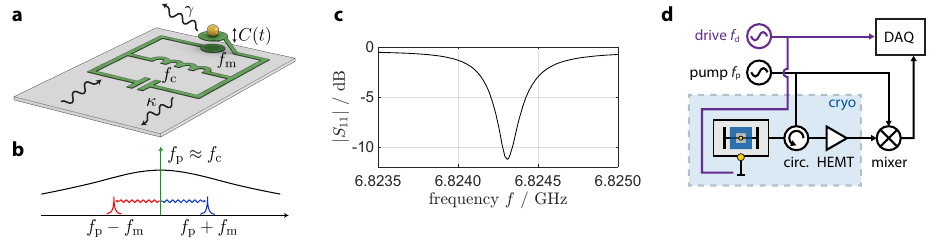}
    \caption{\textbf{Microwave optomechanical measurement.} \textbf a: The microwave mode in the 3D cavity corresponds to an LC-circuit whose frequency $f_\m c$ is modulated by the position of the membrane modeled as a time-dependent parallel capacitance $C(t)$. The membrane, incorporating the test mass, forms the mechanical resonator with the frequency $f_\m m$ and losses $\gamma$. The cavity loses energy at the rate $\kappa$.
    \textbf b: Spectral view of the microwave tone frequencies, which all fall within the cavity response (solid line, not on-scale) in the present unresolved-sideband limit $\kappa/(2\pi) \gg f_\m m$. For readout, the cavity is driven with the  microwave pump tone at the frequency $f_\m p$, and the mechanical resonator imparts sidebands at $f_\m p \pm f_\m m$ on the reflected signal.
    \textbf c: Microwave reflection characterization of the superconducting cavity. The pump tone on cavity resonance is masked for clarity.
    \textbf d: Schematic of the homodyne detection of the mechanical signal. The detection setup is built for coherent acquisition of the gravitational driving (frequency $f_\m d \simeq f_\m m$) for extended periods of time.
    }
    \label{fig:optomechanics}
\end{figure*}

The source mass $m_\m s$ is a gold sphere with a radius of $r_\m s  = 1.02\,\m{mm}$ and is thus comparable to the source mass in Ref.~\cite{westphal_measurement_2021}. As shown in \fref{fig:experiment}\textbf{a}, it is glued to a carbon fiber pole which is actuated up and down with a piezo buzzer disk at the amplitude $z_\m s \simeq 2.5\,\m{\textmu m}$. 
The source mass is mounted via three-axis cryogenic nanopositioners which allow for up to $5\,\m{mm}$ of movement along all axes with sub-micron precision. This way we can adjust the center-of-mass distance of the spheres, and the angle of the dynamic gravitational force.

In spite of the complicated cryogenic setup involving several electromagnetic shielding layers between the source mass and the test mass, they can be brought to a relatively close surface-to-surface distance. 
To ensure no contact between the components even in case of environmental events, we require some minimal gaps of the shield-cavity surface distance $d_\m{c-F}\simeq 40\,\m{\textmu m}$, and also the source mass-shield distance $d_\m{F-s} \simeq 40\,\m{\textmu m}$.
As a result, the masses in our experiment have a minimal surface-surface distance of $d_\m{ss} = d_\m c + d_\m F + d_\m{t-c} + d_\m{c-F} + d_\m{F-s} > 270 \,\m{\textmu m}$, and the gravitationally relevant $d_\m{com,0} = d_\m{ss} + r_\m t + r_\m s > 1.54\,\m{mm}$ (see \fref{fig:experiment}\textbf{b} for the meaning of the dimensions).
At closest approach, we expect a dynamic force of $F_\m G \approx 10\,\m{aN}$.

\textbf{Shielding against vibrations and electromagnetic fields.} Both masses are suspended from the mixing chamber of the dilution refrigerator in their individual acoustic filters (\fref{fig:experiment}\textbf{c}), which comprise of $1.1\,\m{kg}$ copper cubes connected by titanium ring springs. They act as a vibrational low-pass filter, with a cut-off frequency of $\sim 10\,\m{Hz}$ and an attenuation of $\sim 40\,\m{dB}$ per mass-spring block at $2\,\m{kHz}$. The filtering prevents force noise of the pulse tube cryocooler from actuating the test mass oscillator. Additionally, the long chain of filters separating the masses eliminates direct acoustic coupling from piezo to the test mass oscillator.
%
%

Suppressing the effects of electromagnetic forces that would overshadow the weak gravitational force is of utmost importance. A number of mechanisms, including electrical or magnetic coupling, vibrations, and indirect inertial actuation are thinkable.
Direct electrical coupling is relatively easily eliminated by sealing the test mass in the superconducting aluminum 3D cavity.
To prevent more intricate coupling mechanisms, a Faraday shield consisting of two layers of aluminum foil sits in-between driving and sensing stage (see \fref{fig:experiment}\textbf{b}).
Radiated magnetic fields from the source mass drive are effectively canceled by using an electrically floating coaxial drive cable, with the return current being carried by the outer conductor up to the still level of the refrigerator. 
Background magnetic fields are minimized by using only non-magnetic materials, and suppressing Earth's magnetic field in several layers of static magnetic shielding by $\sim 70\,\m{dB}$ down to $\sim 15\,\m{nT}$. For details, see Methods.

\begin{figure*}[t]
    \centering
    \includegraphics[width=0.95\textwidth]{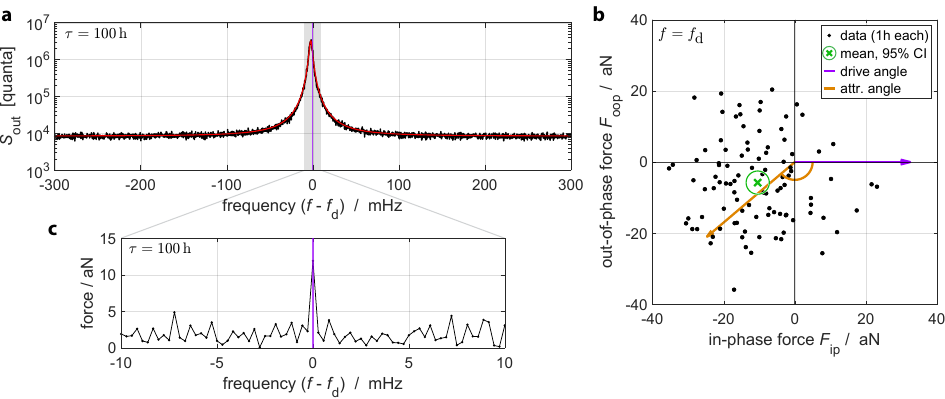}
    \caption{\textbf{Test mass oscillator and gravity signal.}
    \textbf a: Example of a homodyne-detected output spectrum, showing the mechanical thermal noise spectrum (black) appearing as a peak at  $f_\m m = 2138.863\,\m{Hz}$. The spectral density is shown in units of pump frequency quanta at the cavity output reference plane. 
    The red line is a Lorentzian fit. The gravitational drive tone frequency of $f_\m d = 2138.865\,\m{Hz}$ (purple) is close to resonance.
    \textbf b: When considering both the amplitude and phase of data at the driving frequency, individual $1\,\m h$ segments (black points) appear as a cloud in the $IQ$ plane with a non-zero mean (green cross). The expected angle for an attractive force (orange) differs from the angle of the recorded driving tone (purple), due to the phase shift incurred by the near-resonant driving of the mechanical mode.
    \textbf c: In a zoomed section of the spectrum, the excess force at the driving frequency becomes visible when normalizing by the mechanical susceptibility. For this measurement, the spheres are at a close separation of $d_\m{ss} = 0.29\,\m{mm}$.
    }
    \label{fig:dataEval}
\end{figure*}

\textbf{Characterization.} Our microwave optomechanical device (see Fig.~\ref{fig:optomechanics}\textbf{a}) is understood as an LC circuit to which the position of the test mass is coupled parametrically. The coupling is modeled with the optomechanical Hamiltonian $H = \hbar g \LL(a b^\dg + a^\dg b + ab + a^\dg b^\dg \RR)$. Here, $g$ is the effective coupling, $a$ and $a^\dg$ are the ladder operators of the cavity, and $b$ and $b^\dg$ are those of the oscillator. The strength of the interaction is characterized by the cooperativity $C=4g^2/(\kappa \gamma)$. Due to a very small sideband resolution $f_\m m \ll \kappa/(2\pi)$, the counter-rotating terms such as $ab$ are significant. For reading out or controlling the oscillator and thereby the test mass, the cavity is irradiated by a strong microwave pump tone at the frequency $f_\m p$ near the cavity resonance, \fref{fig:optomechanics}\textbf{b}. The cavity is experimentally characterized by a vector network analyzer measurement as shown in \fref{fig:optomechanics}\textbf{c}. The resonance occurs at $f_\m c = 6.8243\,\m{GHz}$, with the total decay rate $\kappa/(2\pi) \simeq 340\,\m{kHz}$. 

The mechanical vibrations of the test mass oscillator are detected in a standard microwave optomechanical detection. Mechanical motion produces sidebands on the reflected pump in the output spectrum of the cavity, which are extracted via homodyne detection at room temperature, \fref{fig:optomechanics}\textbf{d}. From a typical Lorentzian spectral profile as shown in \fref{fig:dataEval}\textbf{a}, we measure the resonance at $f_\m m = 2138.863\,\m{Hz}$, loss $\gamma = 2\pi\cdot 3.67\,\m{mHz}$, and resulting quality factor of $Q_\m m = 5.8 \cdot 10^5$.

For a faithful gravity measurement, the force response of the optomechanical device has to be calibrated using a known force applied to the test mass oscillator. 
Here, we use the separately calibrated microwave noise temperature of the system, which is set by the cryogenic amplifier. This is expressed as white noise spectral density in units of microwave quanta at the cavity frequency, $n_\m{add} \simeq 21.9$. Additionally, the cooperativity is calibrated in a pump+probe measurement with the result $C \simeq 0.35 \pm 0.01$. Using a basic theoretical model for the output spectrum, we can convert the detected spectrum into absolute units at the sample plane, in particular to the thermal noise force and consequently to the gravitational force. We also obtain the mechanical mode temperature $T_\m m \simeq 367 \pm 20\,\m{mK}$ which is slightly above the cryostat temperature $T \simeq 260\,\m{mK}$. The latter is limited by piezo heating $\sim 2\,\m{mW}$.

\textbf{Gravity measurements.} We drive the source mass at the frequency $f_\m d \approx f_\m m$ and amplitude $z_\m s$, which is separately calibrated by optical interferometry at cryogenic temperatures. For readout, we use the homodyne scheme shown in Fig.~\ref{fig:optomechanics}\textbf d. We digitize continuous 1 hour long time traces and acquire $N \geq 20$ time traces per measurement. In parallel, a second digitizer channel records the driving signal fed to the source stage (purple in Fig.~\ref{fig:optomechanics}\textbf d), to keep track of the driving phase. The time traces are Fourier transformed for each measurement. The energy at each Fourier bin, after calibration, is a measure for the force at that frequency.

\begin{figure*}[t]
    \centering
    \includegraphics[width=0.95\textwidth]{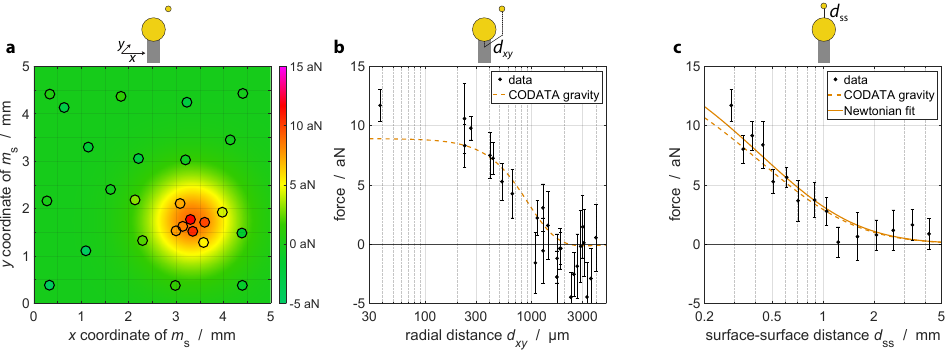}
    \caption{\textbf{Spatial dependence of the measured force.}
    \textbf a: Measured forces at small $z$-separation and different $xy$ source mass coordinates. Data (circles) are well described by a single `hotspot' southeast of the center, whose location we fit to be ($3.3\,\m{mm}$, $1.8\,\m{mm}$). Based on this fit, the theoretically expected gravitational force appears as a color map in the background. 
    \textbf b: The same data (black) appear as a function of radial ($xy$) distance from the center of the hotspot, overlaid by the expected gravitational force in orange. 
    \textbf c: Measured forces at different vertical separation at the center of the hotspot. The solid line is a Newtonian fit. 
    }
    \label{fig:results}
\end{figure*}

Since we also obtain the phases from the Fourier transform, we evaluate and coherently average the full collection of individual time traces $i$. In Fig.~\ref{fig:dataEval}\textbf b, we plot the combination of the amplitudes $F_i$ and phases $\phi_i$ as points in the in-phase (I) and out-of-phase (Q) plane. Their complex mean $F = 1/N \sum_i F_ie^{i\phi_i}$ is marked as a cross.
A phase shift from the driving tone's phase (purple) is expected due to the mechanical susceptibility $\chi_\m m(\omega) = 1/(m_\m t (\omega_\m m^2 - \omega^2 - i\gamma\,\omega))$, which causes the phase shift shown in orange. Within statistical uncertainty, the calculated mean of the data lies on this axis and we conclude that this measurement is consistent with an attractive force between the spheres. Over all our data, we observe no systematic deviation from this `attractive axis' (estimated off-axis force of  $(0.2\pm0.4)\,\m{aN}$). 
An example of a detected $\sim 10$ aN force appears in \fref{fig:dataEval}\textbf c.

We repeat the measurement at different source mass locations, first across the  horizontal ($x$ and $y$) plane. The vertical gaps $d_\m{c-F}$ and $d_\m{F-s}$ remain minimal. Data of these measurements appear as colored circles in Fig.~\ref{fig:results}\textbf a. 
Across most of the plane, there is no detectable force. We do, however, observe a maximum with a force of about $10\,\m{aN}$ south-east of the center and a drop-off in force with a characteristic length scale of about $1\,\m{mm}$. 
This `hotspot' corresponds to the closest approach of the spheres, with their centers of mass aligned along the $z$-axis. By fitting the $x$ and $y$ location of the hotspot to the data using a Newtonian model (underlaid color map), we find test mass coordinates of ($3.34\,\m{mm}$, $1.75\,\m{mm}$). 
It is not surprising that the test mass lies off-center, since sub-millimeter initial alignment of the spheres is not feasible. The same data is shown in Fig.~\ref{fig:results}\textbf b as a function of distance from the hotspot center, with the dashed orange curve representing the expected gravitational force calculated using the gravitational constant given in literature (CODATA, see \cite{mohr_codata_2025}).
Our data match the expected force, when considering the displayed $1\sigma$-confidence intervals.

In a second set of measurements, we move the source mass to the center of the hotspot and measure the force at different vertical separations by changing the $z$-position of the source mass. Data appear in Fig.~\ref{fig:results}\textbf c as a function of the surface-surface distance $d_\m{ss}$ between the spheres. 
We observe a gravitational constant of 
$G_\m{fit} = (7.26\pm0.43\,\m{(stat)}^{+0.73}_{-0.70}\,\m{(syst)})\cdot10^{-11}\,\m m^3\m{kg}^{-1}\m s^{-2}$ (solid line), which
matches the literature value of 
$G_\m{CODATA} = 6.67\cdot10^{-11}\,\m m^3\m{kg}^{-1}\m s^{-2}$ (dashed line)
within the $1\sigma$ systematic uncertainties (see Supplement).
The fact that our data match the expected $d_\m{com}^{-3}$ distance dependence across the entire measurement range is additional indication that we are observing a purely gravitational force: Any force transmitted by vibrations through the suspension system would not depend on $d_\m{ss}$ and data would show a constant offset. Electromagnetic interactions are of surface-surface type, which would manifest as a steep force increase at low $d_\m{ss}$, see Eqs.~\eqref{eq:FcoulombCap} and \eqref{eq:FmovingBoundary}.

\textbf{Discussion.} 
The test mass is subject to a gravitational acceleration of amplitude $a = \frac{2G m_\m s }{d_\m{com,0}^3} z_\m s$ 
independent of $m_\m t$. Detection is limited by the thermal Langevin force noise in the test mass oscillator with the spectral density 
$S_\m f = 4 \gamma m_\m t k_\m B T$, 
and the acceleration noise spectral density 
$S_\m g = \frac{4  k_\m B T \omega_\m m}{Q m_\m t}$. 
The measurement time required to confirm the signal at one standard deviation confidence in coherent averaging is given by 
$\tau =  \frac{S_\m g}{a^2 } = \frac{k_\m B T \omega_\m m d_\m{com,0}^6}{G^2 Q z_\m s^2  m_\m t m_\m s^2 }$, 
scaling very unfavorably with the masses. 
The product $m_\m t m_\m s^2$ may serve as a performance metric when considering the challenge of gravity measurements when both source mass and test mass are scaled down. As compared to Ref.~\cite{westphal_measurement_2021}, this value is here improved by approximately two orders of magnitude, primarily due to the lightness of the test mass. 

The currently best gravity sensors, superconducting gravimeters or cold-atom gravimeters, reach similar or better noise levels on the order nGal$/\rtHz$, or $50\,\m{nGal}/\sqrt{\m{Hz}}$ ($1\,\m{Gal} = 1\,\m{cm}/\m s^2$), respectively. However, the values have been reported in the mHz range, strongly contrasting our kHz case where merely the high frequency gives five orders of magnitude handicap for $\tau$. In the kilohertz regime, work has focused on optomechanical accelerometers, which have reached thermally limited sensitivity of $3\,\m{\textmu Gal}/\rtHz$ at $10\,\m{kHz}$ \cite{Taylor2014Gravimeter} at room temperature, which to our knowledge is the best performance clearly above $1\,\m{Hz}$. In the present work, we reach the gravitational sensitivity  $\sqrt{S_\m g} \sim 6.1 \cdot 10^{-10}\,\m m/\m s^2/\rtHz = 61\,\m{nGal}/\rtHz$, which improves the state-of-the-art by $35\,\m{dB}$. This extraordinary sensitivity allows us to detect gravity due to position changes in the micron range between the masses.

Another interesting benchmark is LIGO. Although not an accelerometer, we can convert the strain sensitivity to a sensitivity towards acceleration of a mirror. At $2\,\m{kHz}$, this gives $\sqrt{S_\m g} \sim$ nGal$/\rtHz$, which  should be reachable in our setup by improving mechanical coherence either by lower mode temperature or higher $Q$. Notice that although our system is intrinsically narrow-band, the bandwidth can be easily extended by orders of magnitude by dissipative optomechanical cooling without compromising the thermally limited sensitivity.

For the exploration of gravity of light objects, the prospects to scale down the masses are excellent. 
Let both masses equal the Planck mass of $22\,\m{\textmu g}$, which corresponds to the energy scale where quantum mechanics and gravity can become equally strong. 
With the following improvements which are likely realizable, we expect similar averaging times as in the current work: source mass motion amplitude $z_\m s \sim 50\,\m{\textmu m}$, surface-surface distance of $100\,\m{\textmu m}$, and mechanical coherence improved by two orders of magnitude.

\textbf{Summary.}
We have demonstrated gravitational coupling between masses of $87\,\m{mg}$ and $1.3\,\m{mg}$ in a new gravity experiment that is enabled by a quantum-ready environment: Our experiment operates at mK temperatures, employs optomechanically controlled micromechanical oscillators
and comparatively high frequencies.
While in the current experiment, the test mass oscillator is limited to a large thermal phonon occupancy at $n_\m m^T \simeq 3.6 \cdot 10^{6}$, we expect that a lower bath temperature and feedback cooling will allow us to approach the ground state. Gravitationally coupling the test mass oscillator to a second mass-loaded membrane \cite{liu_gravitational_2021,Joe2026actuation} is also in sight. Therefore we do not foresee unbearable technological obstacles for experimental realization of non-classical macroscopic mechanical states which would exhibit gravitational coupling. This is not yet true quantum gravity, which still lies years ahead experimentally, but instead \emph{NISG - Noisy Intermediate Scale Quantum Gravity} in analogy to the endeavor towards a universal quantum computer.



\textbf{Supplementary Information} is available in the online version of the paper.

\textbf{Acknowledgements.} We are grateful to Lukáš Flajšman and Jonas Tjepkema for laboratory innovations, and Laure Mercier de L\'{e}pinay for fruitful discussions.
We acknowledge the facilities and technical support of Otaniemi research infrastructure for Micro and Nanotechnologies (OtaNano). This work was supported by the European Research Council (contract 101019712), by European Union’s Research and Innovation Programme Horizon Europe under the Marie Skłodowska-Curie Grant Agreement No.~101198933 (mGramm), and by the Research Council of Finland through the Finnish Quantum Flagship project (project number 358877, Aalto University).




\setcounter{figure}{0}
\renewcommand{\thefigure}{Extended Data \arabic{figure}}

\section{Methods}
\label{sec:methods}
\vspace{-3mm}

\subsection{Microwave-optomechanical device}

\begin{figure}[h]
    \centering
    \includegraphics[width=0.8\linewidth]{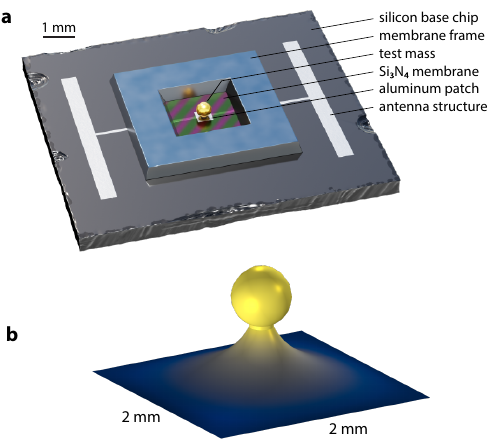}
    \caption{    
    \textbf{a}: Digital render of the sphere-membrane-chip assembly. An aluminum antenna structure on the base chip focuses the electromagnetic mode near the center of the membrane. 
    \textbf{b}: COMSOL simulation of the lowest mode profile of the $\m{Si}_3\m N_4$ membrane loaded with the $1.3\,\m{mg}$ test mass.
    }
    \label{fig:flipchip}
\end{figure}

\vspace{-3mm}
The antenna pattern on the base chip is made of $100\,\m{nm}$ thick aluminum on a $500\,\m{\textmu m}$ thick high-resistivity silicon substrate in an etching process using standard photolithographic techniques. We dice the substrate into a $10\,\m{mm}\times9\,\m{mm}$ chip.
The $\m{Si}_3\m N_4$ membrane -- sold by Silson Ltd
-- is square with dimensions of $2\,\m{mm}\times 2\,\m{mm}$, a thickness of $100\,\m{nm}$, and a prestress of $900\,\m{MPa}$. It sits in a $5\,\m{mm}\times 5\,\m{mm}\times 381\,\m{\textmu m}$ silicon frame. We glue the membrane frame to the center of the base chip. This forms a vacuum gap of $\sim\!1\,\m{\textmu m}$ between membrane and chip, defined by small particles trapped under the membrane frame. A clearly visible interference pattern allows us to verify the distance.
Using electron-beam evaporation in conjunction with a shadow mask, we deposit a $0.5\,\m{mm}\times 0.5\,\m{mm}\times 50\,\m{nm}$ aluminum square in the center of the membrane. Onto this patch, we glue the test mass, creating the device shown in Fig.~\ref{fig:flipchip}\textbf a. Finally, the entire device is glued into an aluminum 3D cavity with the membrane chip facing the cavity wall. For all gluing steps, we use two-component quick epoxy -- Bison International, The Netherlands.
The epoxy cures in about 5 minutes, which is fast enough such that excess flow of the glue does not occur. As the test mass oscillator, we use the lowest drum-like mode of the membrane; see a simulation in Fig.~\ref{fig:flipchip}\textbf b.

\subsection{Gravity source stage}
\label{sec:msMovement}
\vspace{-3mm}

\begin{figure*}[t!]
    \centering
    \includegraphics[width=0.9\textwidth]{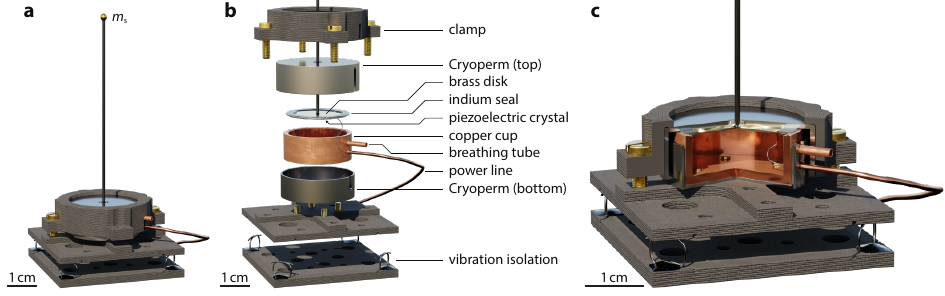}
    \caption{    
    \textbf{Driving stage.} 
    \textbf{a:} The source mass $m_\m s$ connects via a $62\,\m{mm}$ carbon fiber pole to the top of a piezo buzzer.
    \textbf{b:} Exploded view. The stage consists of a number of nested components. The piezoelectric crystal is glued to the bottom of a brass disk. The latter is sealed with indium to a copper cup, with a high-aspect venting hole. Two-part Cryoperm cylinder reduces leakage of AC magnetic fields. A plastic clamp holds everything on top of the copper cup in place. Four small beryllium copper ring springs provide $20\,\m{dB}$ of initial vibration isolation to the stack of nanopositioners below (not pictured).
    \textbf{c:} Cutaway view of the assembly.
    }
    \label{fig:drivingStage}
\end{figure*}

The gravity source mass is connected via a $62\,\m{mm}$ long, $1\,\m{mm}$ wide carbon fiber pole to an electromagnetically shielded piezo disk, see Fig.~\ref{fig:drivingStage}. The disk is taken from a commercial buzzer (Murata PKM22EPPH2001-B0) and consists of a brass disk with the piezoelectric material (likely PZT) on one side. 

The piezo driving voltage and return currents create a potential source of non-gravitational coupling via electromagnetic forces. It is therefore critical to suppress electromagnetic fields from radiating out. The active electrode of the piezo disk is contained in a Faraday cage consisting of the brass disk on top and a copper cup beneath. An indium seal between the two prevents leakage of electric fields, and a high-aspect ratio breathing tube allows for venting. Surrounding the cup and the disk are two $1\,\m{mm}$ thick high-permeability (Cryoperm) cups forming a shell of magnetic shielding, with a circular hole in the upper one giving space to the carbon fiber pole. The magnetic shield is intended to confine AC magnetic fields created by uncancelled piezo currents occurring in the disk region. A clamp made of carbon-reinforced plastic holds the construction in place.

We supply the $65\,\m{V}$ driving voltage to the piezo buzzer via a custom coaxial power line. The outer conductor is soldered to the copper cup, essentially extending the Faraday cage along the power line up to the still platform of the cryostat. This also means that the driving stage components and the power line are electrically floating. The piezo return current flows almost exclusively in the outer shell of the coax, thus canceling magnetic fields radiating out. Incidentally, the floating coax also acts as a thermal link to the gravity source stage, dumping some of the large heat load from the vibrating disk to the still platform, which has a higher cooling power than the mixing chamber platform. Nevertheless, the heat load from the piezo limits our source mass amplitude.

\subsection{Calibrations}
\vspace{-3mm}

\subsubsection{Cooperativity}
\vspace{-3mm}

\begin{figure}[t]
    \centering
    \includegraphics[width=0.99\linewidth]{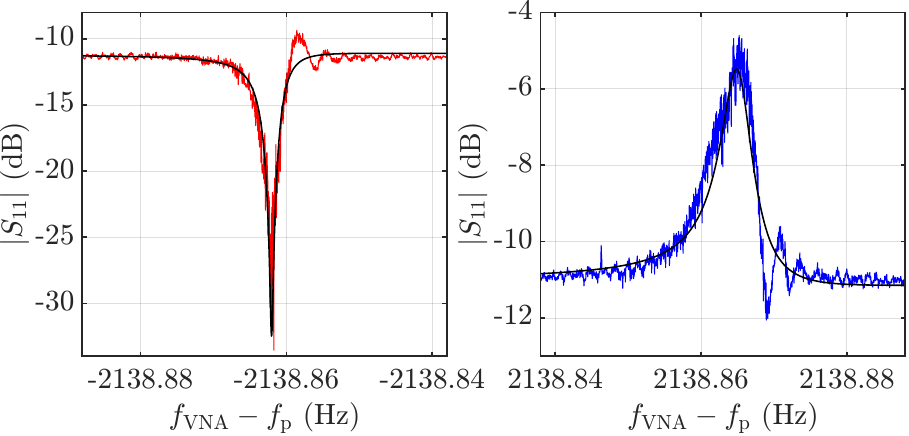}
    \caption{\textbf{Cooperativity calibration.} 
     Optomechanically induced transparency measurement around the red (left) and blue (right) sidebands, as measured with the VNA. The solid lines are fits according to theory. There are slight offsets in the peak positions because of the VNA operation in the exceptionally narrow frequency range.}
    \label{fig:OMIT}
\end{figure}

The effective optomechanical cooperativity is calibrated via a pump and probe measurement (OMIT) \cite{Kippenberg2010OMIT}. The OMIT response serves as a probe which is sensitive to the cooperativity and possible pump detuning. This is carried out under the same conditions, including the pumping, as the gravity measurements. We probe the linear response both at the red and blue mechanical sidebands with a Vector Network Analyzer (VNA). The VNA probe frequency sweep has to be carried out much slower than $\gamma^{-1}$ to equalize the driven oscillator to a steady state at each point. The two sidebands, as seen in the data in  \fref{fig:OMIT}, are different due to  opposite interference that becomes apparent in the non-resolved sideband regime. The small ringing, which we ignore, is due to the oscillator being slightly excited by preceding frequency points in the left-to-right sweeps.

We calculate the theoretically expected response at arbitrary sideband resolution. Let $\Delta= \omega_\m p - \omega_\m c$ be the pump detuning from the cavity. When $\Delta=0$ and in the unresolved-sideband approximation, the response at the two sidebands is given by
\begin{equation}
\begin{split}
S_{11} = 1 - \frac{2 \kappa_e}{\kappa} - \frac{2 \kappa_e}{\kappa} \frac{ i  C \gamma \omega_m}{ (\gamma/2 - i \omega)^2 + \omega_m^2} \,.
\end{split}
\end{equation}
The numerical fitting uses the full expression from the non-rotating-wave Hamiltonian. Here, $\kappa_e/(2\pi) \simeq 120\,\m{kHz}$ is the external coupling rate of the cavity.

Beside cooperativity, we also obtain a negligibly small pump detuning $\Delta/(2\pi) = 5\,\m{kHz}\pm 0.4\,\m{kHz}$.  Such a small detuning has no effect on the output spectrum, and thus for simplicity we set the detuning to zero when discussing the spectra.

\subsubsection{Noise calibration}
\vspace{-3mm}

Microwave opto/electromechanics experiments carried out at millikelvin temperatures involve stages of amplifications and attenuating microwave cabling. Recovering the value of total gain from the sample output plane to volts displayed at the computer is not straightforward. Additionally, one has to consider the transfer function of the optomechanical device itself; what force applied to the sample would produce a given output signal.

Typically, a calibration is carried out via a temperature sweep or sideband asymmetry, allowing one to recover the thermal Langevin force driving the oscillator. However, these are not available to us. A thermal sweep requires access to an order of magnitude range in temperature, while our range is restricted from below by thermalization, and from above by the superconducting transition of aluminum.

We thus choose to use the microwave noise level of the detection system as the known signal, calibrated at the sample plane. The noise calibrations were performed in a separate cooldown because introducing any additional components may compromise the magnetic environment or return current paths. We used the standard technique: A $50\,\Omega$ resistor is mounted to a heater and a RuO$_2$ thermometer, and connected in place of the electromechanical sample. 
The heated resistor emits Johnson noise at the adjustable, known temperatures $T_{\m{in}}$. At the room temperature detection, the spectral density is $S_\m{d} = S_{\m{in}} \alpha \mathcal{G} + S_\m{HEMT} \mathcal{G}$. Here, the Johnson noise spectral density is $S_{\m{in}} = k_\m B T_\m{in}$, $\alpha < 1$ is the transmission between the sample and the cryogenic microwave amplifier (HEMT), $\mathcal{G}$ is the system gain, $S_{\m{HEMT}} = k_\m B T_\m{HEMT}$ is the HEMT noise, and $T_{\m{HEMT}}$ is the HEMT noise temperature. By measuring the noise power as a function of $T_\m{in}$, we obtain the effective noise from a linear fit: $S_{\m{HEMT}}/\alpha \equiv S_{\m{add}}$. We can express this as the effective system noise temperature $T_{\m{add}} = S_{\m{add}}/k_\m B$.

We primarily work with dimensionless spectral density, in  units of `quanta'. This is the natural unit system in optomechanics theory. In these units, the input noise is $n_\m{in} =  S_\m{in}/(\hbar \omega)$, and the system noise in quanta is $n_\m{add} =  S_\m{add}/(\hbar \omega)$. 

\begin{figure}[t]
    \centering
    \includegraphics[width=0.6\linewidth]{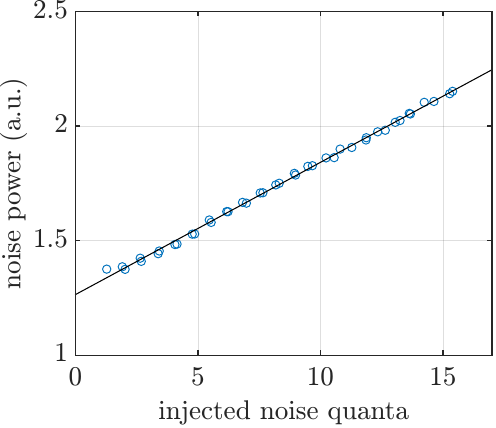}
    \caption{\textbf{Noise calibration.} 
     Detected noise level as a function of temperature of the $50\,\Omega$ noise source. The operation frequency is the cavity frequency $f_\m c$.}
    \label{fig:noisecalib}
\end{figure}

We show in \fref{fig:noisecalib} an example of the system noise calibration and the linear fit. The resistor temperature is varied between $0.4\,\m K$ and $5.0\,\m K$.
We obtain $n_\m{add} \simeq 21.9$ (corresponding to $T_\m{add} \simeq 7.18\,\m K$) with negligible statistical errors. The value for added noise is  typical for our cryostats running similar amplifiers.

The added noise would be expected to equal the measured noise floor of the mechanical thermal peak. However, as seen in \fref{fig:dataEval}\textbf{a}, the noise background is much larger than $n_\m{add}$. This is due to vibrations in the setup from an unverified source, which optomechanically couple to the cavity and cause broadband noise background. The calibration for true noise background equal to $n_\m{add}$ is found by switching the pump tone off, in which case the optomechanical background disappears. The other possibility is to switch the noise calibration frequency off from the cavity by an amount $\gg \kappa$, but we prefer the former method since $n_\m{add}$ is frequency dependent.

\subsubsection{Mode temperature calibration}
\vspace{-3mm}

For the optomechanics theoretical treatment below, we ignore backaction in any form, as it is negligible with our parameters at cooperativity $C < 1$, as well as cavity heating. The homodyne-detected  (single-sided, $\omega > 0$) output noise spectrum referenced at the sample plane is in the unresolved-sideband case
\begin{equation}
\begin{split}
& S_\m{out}(\omega) = \frac{g^2 \kappa_\m e \kappa^2}{\LL[(\kappa/2)^2 + \Delta^2\RR]^2} S_{x}(\omega) + n_\m{add}\,,
\end{split}
\end{equation}
with the mechanical spectrum for $\omega > 0$
\begin{equation}
\label{eq:Sx}
\begin{split}
&  S_{x}(\omega) = \frac{\gamma}{2} \frac{n_\m m^T + \puoli}{(\gamma/2)^2 + (\omega-\omega_m)^2}\,.
\end{split}
\end{equation}
For the noise spectrum, we can set the pump detuning $\Delta = 0$, which is satisfied well enough in the experiment. The output spectrum is
%
\begin{equation}
\label{eq:SoutFinal}
\begin{split}
& S_\m{out}(\omega) = 4 C \gamma \frac{\kappa_\m e}{\kappa} S_{x}(\omega) + n_\m{add} \,.
\end{split}
\end{equation}
At room temperature after detection, the detected spectrum is
\begin{equation}
\label{eq:Sd-gain-Sout}
\begin{split}
& S_\m{d}(\omega) = \mathcal{G} S_\m{out}(\omega) \,.
\end{split}
\end{equation}

The detected noise background with pump off is $S_\m{d,off} = \mathcal{G} n_\m{add}$, which calibrates the system gain $\mathcal{G}$ and allows us to convert a detected spectrum $S_\m{d}$ into the sample plane via \eref{eq:Sd-gain-Sout}. The mechanics mode temperature $T_\m{m}$ is derived by finding the integral of the Lorentzian fit to the thermal peak, inverting \eref{eq:SoutFinal}, and integrating and inverting \eref{eq:Sx}. The thermal energy of the oscillator is $E^\m{th} = k_\m B T_\m m$.

\subsubsection{Force calibration}
\vspace{-3mm}

During gravity measurements, we acquire a set of $N$ time traces $V_i(t)$ with a DAQ. In order to convert these into a force acting on the test mass, we first perform fast Fourier transforms (fft) to obtain the power spectral densities $S_i(\omega)$, with discrete frequency bins of width $\Delta\omega$. We fit a Lorentzian $\mathcal L(\omega)$ to the average $1/N\sum_i S_i$ of these spectra to find the linewidth $\gamma$ and amplitude $c_0$. 
The integral of the Lorentzian
$E_\m L = \frac{\pi}{2}\,\gamma\,c_0$ 
(in units of $\m V^2$) is a measure for the energy of the mechanical oscillator, which is dominated by its (rms) thermal motion 
$x_{\mathrm{th}} = \sqrt{k_\m B T_\m m / (m_\m t\,\omega_\m m^2)} = 150\,\m{fm}$.
Dividing these quantities thus gives us the conversion factor 
$\alpha = \sqrt{E_L}/(\sqrt 2 \, x_\m{th})$ 
in units of V/m, where we need to add the factor $\sqrt 2$ to move from rms motion to amplitude. The mechanical susceptibility 
$\chi_\m m(\omega) = 1/(m_\m t \cdot (\omega_\m m^2 - \omega^2 - i\gamma\,\omega))$
then allows us to convert from length to force units.
We thus obtain the force exerted on the force sensor at each frequency bin $\omega$ of the fft during a single time trace via 
\begin{align}
    \label{eq:forceFromSpectrum}
    F_i(\omega) &= \frac{\sqrt{S_i(\omega)  \cdot \Delta \omega}} 
    {\alpha \cdot \left|\chi_\m m(\omega)\right|} \\
    &= \sqrt{\frac
    {4\,k_\m B T_\m m\,S_i(\omega) \Delta \omega}
    {\pi\gamma c_0 m_\m t \omega_\m m^2\left|\chi_\m m(\omega)\right|^2}
    }    
    \,. 
\end{align}
We average the forces at the driving bin $\omega_\m d$ as depicted in \fref{fig:dataEval}\textbf b, and finally project the resulting complex force vector onto the attractive axis to obtain a final force in newton.

\subsubsection{Distance calibration}
\vspace{-3mm}

During cooldown, thermal contraction changes the relative positions of the masses and of the Faraday shield on the order of millimeters. Thus the cavity-Faraday shield distance $d_\m{c-F}$ and Faraday shield-source mass distance $d_\m{F-s}$, as well as the lateral $x$, $y$ offset of the masses have to be calibrated \textit{in situ}. We use the nanopositioners with vertical movement to purposefully collide the components, which is visible as a strong impulse signal in the test mass oscillator. This gives us micron-level accuracy for the calibration of the vertical distances, which stay fixed within ten microns for months. We occasionally repeat the distance calibrations. For the test mass position inside the 3D cavity we do not have such a direct calibration tool, and it can not be observed visually since it must be inside a closed conducting enclosure. Thus, alignment of the spheres along the $x$ and $y$-directions occurs via locating the `hotspot'.
\\

\subsubsection{Source mass motion calibration}
\vspace{-3mm}

To determine the source mass movement amplitude $z_\m s$, we perform a direct optical measurement at millikelvin temperatures. For this, we must replace the standard cryostat cans with windowed cans, which is why the calibration is done in a separate cooldown. We measure the vibration amplitude of the source mass via a Doppler-shift vibrometer. For our standard driving frequency,
we find $z_\m s = (2.5\pm0.1)\,\m{\textmu m}$ at a driving amplitude of $65\,\m V$. By repeating the measurement along slightly different optical axes, we estimate the tilt of the source mass movement axis with respect to vertical to be on the order of $0.5^\circ$. Furthermore, we are able to measure the full amplitude spectrum and find the base resonant mode of the gravity source stage to be the base drum-like mode of the piezo disk at 3.71 kHz with the quality factor of 370.
We are thus far off-resonant. There are no further relevant modes in the gravity source stage, including bending modes of the carbon fiber pole.
\\

\subsection{Test mass and source mass geometry}
\vspace{-3mm}

We use gold for both masses due to its high density ($19.3\,\m g / \m{cm}^3$), high chemical stability, magnetic purity, and commercial availability. 
The spherical geometry of the masses was chosen for simplicity. While more complicated geometries can significantly enhance the dynamic gravity for given masses \cite{Liu2026GravityFormFactor}, the added complexity does not justify this step currently.

The test mass is a 4N purity gold sphere with a radius of $250\,\m{\textmu m}$, purchased from Goodfellow Advanced Materials.
It is highly spherical, with a diameter variation of $\lesssim~1\,\%$. We calculate the mass based on the radius to be $m_\m{t}=(1.26\pm0.02)\,\m{mg}$.
We produce the source mass in-house by melting multiple 4N purity gold nuggets together and then etching down to the desired mass. As a result, the source mass is less spherical than the test mass, with vertical and horizontal axes differing by about $3\,\%$. We weigh the source mass directly to be $m_\m s = (86.6\,\pm0.1) \,\m{mg}$.

\subsection{Non-gravitational coupling channels}
\vspace{-3mm}

We estimate the forces produced by the identified nongravitational coupling mechanisms and compare the expected responses with the measured gravitational signal. The estimates use the mechanical transfer function \eref{eq:HmN} below and the electrical model in the Supplement. We aim at order-of-magnitude estimates because the transfer function value may contain significant systematic uncertainties.  Where possible, we test the estimates by deliberately driving the relevant component.


We believe that the strongest potential cross-coupling channel is not direct driving of the test mass by electromagnetic fields, but actuation of the vibrational filter suspension, which then inertially actuates the test mass oscillator.
This is a highly sensitive mechanism  to excite the test mass oscillator, because any oscillation of the suspension in the vertical direction turns into a $Q$ times larger amplitude in the oscillator.

Direct acoustic coupling is analyzed in the Supplement, and found to be negligible based on modeling and experiment.

\subsubsection{Contact potentials}
\vspace{-3mm}

The electrostatic force across a capacitor $C(z)$ with a vacuum gap $z$ and a voltage $V(t)$ is
\begin{equation}
\begin{split}
\label{eq:FcoulombCap}
F(t) = \frac{1}{2}C'(z) V(t)^2 \,.
\end{split}
\end{equation}
In order to create a force at the driving frequency $\omega_\m d$ in this capacitor, a constant DC electric field on top of an AC component at $\omega_\m d$ is necessary. While such a DC component could be caused by trapped charges, a stronger mechanism is the difference in contact potential. The work function of a given metal can be strongly influenced by surface morphology or preparation. Microscopically, the local variation of work function is called patch potentials. Thus, although the conductive parts are galvanically connected at the same electrochemical potential, an electric field may be present. Between different macroscopic surfaces made of nominally the same material, the potential difference may be significant. In the following, we assume the DC voltage $100\,\m{mV}$ between the considered metal surfaces, which is a worst-case estimate.

\subsubsection{Direct electrical coupling}
\vspace{-3mm}

Consider a test mass oscillator with full exposure to the electric fields from the electrodes in the source piezo ca. $65\,\m{mm}$ away. A crude estimate of the capacitance between the piezo disk and the test mass is $C(z) \sim 10^{-14}$ F, and the resulting force driving the oscillator is $F \sim 10^{-12}\,\m N$. The oscillator amplitude $F/(m_\m t \gamma \omega_\m m)$ becomes so high that the cross-coupling would dominate gravity by $\sim 110\,\m{dB}$. Shielding the test mass from electric fields from the driving stage is therefore necessary, but this task is accomplished simply by embedding the test mass inside the 3D cavity. 

For a quick estimate of the efficiency of the cavity enclosure shielding, consider leakage of electric fields through a long hole. Per length of the hole equal to one diameter, the attenuation is on the order of $40\,\m{dB}$. The SMA pin breathing hole in the cavity wall has an aspect ratio $\sim 5$, which immediately results in more shielding than needed. In reality, the fields are blocked at the source side as well, even more carefully in fact (see description of the driving stage in Supplement). Moreover, there is the `real' Faraday shield in between.


%

\subsubsection{Suspension response}
\vspace{-3mm}

We estimate the inertial actuation of the test mass oscillator using a finite-element model of the suspension, applying a vertical force $F_\m{inert}$ to the lowest cube. The magnitude response of the suspension at $2\,\m{kHz}$ is computed as 
\begin{equation}
\begin{split}
\label{eq:HmN}
H_\m{m/N} = \frac{z_\m{inert}}{F_\m{inert}} \sim 10^{-8} \, \m{m/N} \,,
\end{split}
\end{equation}
and the oscillator's response is
\begin{equation}
\begin{split}
\label{eq:inertresponse}
z_\m{stray} = Q H_\m{m/N} F_\m{inert}\,.
\end{split}
\end{equation}

Thus, in order to compete with a typical weak gravitational driving $z_\m{t} \sim 10\,\m{fm}$, as little as $z_\m{inert} \sim 10^{-20}\,\m m$ suspension oscillation is enough. The corresponding force driving the suspension then becomes $F_\m{inert} \sim 10^{-12}\,\m N$. Below, we estimate mechanisms that would cause such a force to the lower parts of the test mass suspension.

\subsubsection{Vibrations within line-of-sight}
\vspace{-3mm}

As discussed above, the electric fields due to driving are already well blocked by separately enclosing the test mass oscillator, as well as the driving electrodes. The question arises, whether we then need the `real' Faraday shield between the source mass and the cavity wall. The shield is necessary, but not because of further blocking of electric fields from reaching the test mass.

Consider a capacitor where one electrode's position $z(t)$ is time-dependent. The other electrode experiences a force
\begin{equation}
\begin{split}
\label{eq:FmovingBoundary}
F(t) & \approx \frac{1}{2} C''(z) V^2  z(t)\,.
\end{split}
\end{equation}
Thus, a constant field due to the contact potential, and an oscillating boundary, is enough to cause a force at the oscillation frequency.

Let the source mass oscillate below the cavity without the Faraday shield, at a typical close separation. We find based on a sphere-plane capacitance model $C \sim 0.3\,\m{pF}$ and $F \sim 0.2\,\m{pN}$, which now is interpreted as $F = F_\m{inert}$ exciting the suspension. From \eref{eq:inertresponse} we find, $z_\m{stray} \sim 1\,\m{fm}$, which, while smaller than the typical gravity response, would not be acceptable given rough estimates involving systematic uncertainties.

\subsubsection{Faraday shield vibrations}
\vspace{-3mm}

With the Faraday shield in place, the force in \eref{eq:FmovingBoundary} still exists, but it is now exerted on the shield. The force is then transmitted capacitively to the test mass. The Faraday shield and the 3D cavity form a plate capacitor of $\sim 3\,\m{cm}^2$ area. At a typical vacuum gap of $d_\m{c-F} = 50\,\m{\textmu m}$, the capacitance is $C_\m{c-F} \simeq 50$ pF. With these numbers, $F_\m{c-F}$ becomes negligible ($\lesssim 0.1$ pN) if the induced shield vibration amplitude is $z_\m{F} \lesssim 1\,\m{nm}$.

We verified this experimentally: In a calibration cooldown where the cryostat was equipped with optical windows, we studied with Doppler vibrometer how vibrations of the Faraday shield induce actuation of the test  mass oscillator. The shield was set to vibrate by applying the AC driving voltage to the nanopositioner holding the shield. Nanometers of shield vibrations indeed actuated the mechanics via the indirect inertial actuation route. When instead the driving was applied to the source stage as during normal operation, the shield vibrations were undetectable, below $\sim 50$ fm. Based on this, we infer that shield vibrations in a normal run are fully negligible.
\\
Note also that shield vibrations are the enabling method to perform the shield - source mass position calibration. At $m_\m s$-shield contact, the shield vibrates strongly ($\sim\m{\textmu m}$), which, based on our model, causes inertial actuation of the mechanics up to picometer amplitudes, overwhelming even the thermal motion and are thus easily observed in real time.

%
%


\subsubsection{Time-dependent electric field in the shield - cavity gap}
\vspace{-3mm}

Consider \eref{eq:FcoulombCap} with a $100\,\m{mV}$ contact potential difference, plus a small AC component at $\omega_\m d$. We find that an AC voltage of $\m{\textmu V}$ causes an unacceptable $z_\m{stray} \sim 1\,\m{fm}$. Besides the voltage induced by the common-impedance coupling discussed below, we have not been able to imagine a process that would cause such an AC voltage in the shield-cavity gap.

\subsubsection{Charges on the source mass}
\vspace{-3mm}

Let $q_\m{trap}/e = 10^7$ charges (two orders of magnitude more than in Ref.~\cite{westphal_measurement_2021}) reside on the source mass, which is oscillating at a close distance $d$ to either the cavity bottom or Faraday shield. The analysis is similar to above, but in the language of image charges. For a sphere, the total force at a small gap is 
$F_\m{tot}  \simeq q_\m{trap}^2 (4\pi\varepsilon_0 r_\m s d[ \ln(2r_\m s/d)+2\gamma_\m E]^2)^{-1}$,
where $\gamma_\m E\simeq0.58$. The dynamic force is $F \approx F_\m{tot} z_\m s/d$. We find $F \sim 1\,\m{nN}$, which, if driving the suspension directly without Faraday shield in between, would cause overwhelming inertial actuation. As above, adding the shield will solve the issue.

\subsubsection{Charges on the test mass}
\vspace{-3mm}

We expect on the order $10^7$ charges on the test mass or on the electrode beneath it based on typical DC offset measurement \cite{Joe2026actuation}. However, the test mass is strongly shielded by multiple layers of superconducting material, and it is difficult to imagine a process that would cause cross-coupling due to these charges. The exception is a possible small remnant of the AC driving voltage on the SMA pin, induced by common-impedance coupling. This is mentioned below, and concluded unlikely based on modeling. We also excluded this possibility in an experiment, by intentionally applying a $1\,\m V$ signal at $\omega_\m d$ to the microwave lines at room temperature, and observed no coupling. The driving signal hitting the SMA pin here was estimated to be in the $\m{\textmu V}$ range.

\subsubsection{Common-impedance coupling}
\vspace{-3mm}

Let us consider an electrical circuit model of the source piezo driving, paying attention to reactive responses of the components inside the refrigerator (see Supplement). Although all metallic components are galvanically connected, they are not at the same AC potential. This is simply because the piezo return currents cause voltage drops. The circuit model gives rough estimates for voltage differences between components of interest in the setup, or AC currents:

\begin{itemize}
    \item 
The AC voltage between cavity and the shield is on the order $\sim 20\,\m{nV}$. The resulting $F_\m{inert} \sim 1\,\m{fN}$ is negligible by $40\,\m{dB}$.

\item Including in the circuit model the measurement SMA pin (not shown), we obtain $\sim 20\,\m{nV}$ AC between the pin and the test mass. Actuation is negligible. 

\item The AC voltage between still can (not shown) and cavity is on the order tens of $\m{\textmu V}$. Actuation is negligible.

\item The return current between the Faraday shield and the cavity is on the order $0.2\,\m{pA}$, which we consider is negligible.

\end{itemize}

\subsubsection{Magnetic coupling}
\vspace{-3mm}

We treat our bulk aluminum components as type I superconductors. The length scale of Meissner screening of low-frequency magnetic fields is the London penetration length, on the order $100\,\m{nm}$. Thus, if the local state is superconducting (see possible exceptions below), magnetic flux does not penetrate the Faraday shield or the cavity at all.

\subsubsection{Forces due to piezo return currents}
\vspace{-3mm}

In an ideal case, the piezo return currents flow exclusively on the outer conductor of the coaxial cable, and any magnetic field radiation is canceled. Let us make a worst-case estimate, which assumes the canceling is incomplete. Let all the piezo current $\sim 2\,\m{mA}$ flow through the suspension. The Lorentz force is on the order of $0.1\,\m{pN}$, which may be significant for inertial actuation. This situation where a significant parts of the currents would flow through the suspension, however, is fully unrealistic and we conclude that under the experimental conditions, magnetic forces due to piezo currents can be neglected.

\subsubsection{AC currents due to charges on source mass}
\vspace{-3mm}

Above, we discussed how the expected significant amount of charges residing on the moving source mass may cause electrostatic interference. Besides this, also magnetic interference may arise due to currents induced by the moving charges. The AC current flowing in the Faraday shield, circularly symmetric with respect to the source mass position, is on the order of $q_\m{trap} \omega_\m d z_\m s/r_\m s$. The current magnitude grows possibly to a significant amount, on the order $100\,\m{pA}$. However, because of the Meissner screening, these currents flow only near the lower surface of the lower shield foil, and their total magnetic field does not reach the upper side. Notice these currents correspond to local distribution of charges in the shield at the scale of the source mass diameter, and they do not appreciably reach beyond the shield.

\subsubsection{Trapped magnetic flux}
\vspace{-3mm}

We use multiple layers of magnetic shielding to suppress Earth's ambient field to the nT range: At room temperature, a mu-metal can surrounds the cryostat. The still radiation shield is made of Cryoperm. We can estimate that this is sufficient to prevent any magnetic flux being trapped in the experiment. 
The 3D cavity can be modeled as a superconducting box.
For a (worst-case) wall thickness $d_\m c$ and characteristic size $D$, flux does not enter the enclosure below a field on the order of $B_\m c d_\m c/D\approx 30 \,\m{\textmu T}$. This is over three orders of magnitude higher than our estimated ambient magnetic field and the cavity thus serves as a complete Meissner shield.
Regarding the Faraday shield, consider a superconducting plate exposed to a perpendicular magnetic flux. Flux will penetrate the plate in the form of normal stripes in a configuration determined by the free energy of the system. The applied flux is concentrated to stripes associated with the bulk critical field ($B_c \sim 10\,\m{mT}$ for aluminum). With our dimensions, the stripes in the bulk components have a width of $\sim 20\,\m{\textmu m}$, and are spaced to occur every $10\,\m m$. We thus expect them not to occur at all in the Faraday shield.

\subsubsection{Other gravitational interactions}
\vspace{-3mm}

We have considered a number of gravitationally interacting components other than a direct coupling of the two spheres. This includes interactions between the test mass and various parts of the driving stage, and interactions between the source mass and various parts of the sensing stage. All interactions have an effect of $<1\%$ and are listed as systematic uncertainties in the Supplement.


\end{document}